\documentclass[aps,twocolumn,showpacs]{revtex4}
\usepackage{graphicx}   
\usepackage{latexsym}   

\newcommand{\bold}[1]{\mbox{\boldmath $#1$}}    
\newcommand{\ie}{{\em i.e.}}			
\newcommand{\etal}{{\em et al.}}		

\begin{document}

\title{Brownian shape motion
on five-dimensional potential-energy surfaces:\\
Nuclear fission-fragment mass distributions}

\author{J{\o}rgen Randrup$^a$ and Peter M{\"o}ller$^b$}

\affiliation{
$^a$Nuclear Science Division, Lawrence Berkeley National Laboratory,
Berkeley, California 94720, USA\\
$^b$Theoretical Division, Los Alamos National Laboratory, 
Los Alamos, New Mexico 87545, USA}

\date{February 3, 2011}

\begin{abstract}
Although nuclear fission can be understood qualitatively as an evolution
of the nuclear shape, a quantitative description has proven to be very elusive.
In particular, until now, 
there exists no model with demonstrated predictive power
for the fission fragment mass yields. 
Exploiting the expected strongly damped character of nuclear dynamics,
we treat the nuclear shape evolution in analogy with Brownian motion and
perform random walks on five-dimensional fission potential-energy surfaces
which were calculated previously and are the most comprehensive available.
Test applications give good reproduction of
highly variable experimental mass yields.
This novel general approach requires only a single new global parameter,
namely the critical neck size at which the mass split is frozen in, 
and the results are remarkably insensitive to its specific value.
\end{abstract}


\maketitle


Since nuclear fission was discovered in 1938 \cite{Hahn39},
its theoretical modeling has presented significant challenges.
As discussed already in the pioneering papers by 
Meitner and Frisch \cite{Meitner39} 
and Bohr and Wheeler \cite{BohrNat143,BohrPR56} in 1939,
nuclear fission can be viewed qualitatively as an evolution
of the nuclear shape from that of a single compound nucleus
to two receding fragments.
But the character of the shape dynamics is still not well established.
Nor is it yet understood in detail how the original compound nucleus
is transformed into a variety of different fragmentations.
Therefore models for the resulting distribution of mass splits have, 
until now, had limited predictive power.

The currently used methods for calculating fragment yields
include a variety of phenomenological approaches.
These invariably introduce a number of parameters whose 
values are determined from adjustments to measured mass yields
and other observables \cite{BrosaZPA325,BrosaPRep197,BenlliureNPA628}.
Such approaches typically reproduce experimental data
in the regions where their parameters were determined,
but they do not advance our understanding of fission
and they also tend to fail when applied to other regions.

Scission models consider the relative statistical weights of
various fragmentations at the time of scission \cite{FongPR102,WilkinsPRC14}.
Such calculations are entirely static in nature 
so they cannot take account of any dynamical effects.
While scission models often yield reasonable agreement with the observed
mass distributions, they are not universally successful
and their failures suggests that the resulting divisions
are sensitive to the pre-scission dynamics.

A number of dynamical models of fission have also been developed.
Most of these concentrate on the average evolution
and they are often macroscopic.
Langevin treatments have been developed and applied for excitations 
sufficiently high to render the dynamics macroscopic
\cite{KarpovPRC63,NadtochyPRC65,RyabovPRC78}.
The presently most refined microscopic approach
uses the time-dependent generator-coordinate method
with Hartree-Fock-Bogoliubov states 
based on an effective interaction \cite{GouttePRC71,DubrayPRC77}.
This treatment is rather computer intensive and,
consequently, only a few systems have been studied so far.
In particular,
fragment mass distributions have been calculated 
only for fission of $^{238}$U \cite{GouttePRC71}
(though results for $^{240}$Pu are underway \cite{Younes}).

We introduce here a novel method for calculating
fission fragment mass distributions.
It invokes the expected dissipative character of the
coupling between the nuclear surface and the internal degrees of freedom
\cite{BlockiAP113} which, 
in the Smoluchowski limit of strong coupling \cite{Abe},
gives the shape dynamics the character of Brownian motion.
Consequently, once the potential energy is known as a function of deformation
for a sufficiently rich family of fission shapes,
the required calculation is relatively straightforward,
amounting effectively to a random walk 
on the corresponding potential-energy surface.
Because suitable deformation-energy surfaces are available for
essentially all nuclei of potential interest \cite{MollerPRC79},
this treatment of the fission dynamics
provides a powerful predictive tool.
We describe here the key features of the approach 
and present several test applications.


As mentioned above,
our method exploits the limit of strong dissipative coupling
between the nuclear surface motion and the internal degrees of freedom,
as is characteristic of systems dominated by one-body dissipation 
\cite{BlockiAP113}.
The basic mechanism is the reflection of individual nucleons
off the moving surface which generates a dissipative force
that is rather strong due to the nucleonic Fermi motion.
The {average} nuclear shape evolution is then determined by the balance 
of the associated dissipative force on a surface element
and the conservative force due to the deformation energy.
The resulting equations of motion for the shape evolution
contain no adjustable parameters and dynamical fission calculations 
yield remarkably good agreement with data 
for the most probable fragment-kinetic energies
\cite{BlockiAP113}.

Because the individual nucleons reach the moving surface at random times, 
the associated force is stochastic,
in accordance with the fluctuation-dissipation theorem \cite{EinsteinAdP17}.
The present treatment is the first implementation of the stochastic part
of the one-body mechanism for mononuclear dynamics
in the Smoluchowski limit
(it was implemented early on for nucleon exchange
in the dinucleus \cite{RandrupNET} where it
proved to be essential for understanding the
dependence of the mass distribution on energy loss
in damped nuclear reactions \cite{SchroderPRL44}).

In this physical picture,
the evolution of the nuclear shape is akin to Brownian motion.
For the fissioning nucleus, the shape plays the role of the Brownian body,
while the environment consists of the microscopic degrees of freedom
associated with a nucleon gas.
For a given shape, the nuclear surface is being continually
assaulted by those nucleons;
the average of these impulses yields the associated friction force and
the residual fluctuations give the shape evolution a diffusive character.

While it is generally somewhat complicated to treat this dynamical problem,
considerable simplicity emerges in the strongly damped limit
which is expected to be a reasonable starting point 
for the description of nuclear dynamics \cite{RandrupAP125}.
Indeed, the strongly damped {(Smoluchowski)}
 limit of standard Brownian motion
can be treated as a random walk in configuration space
and the overall friction strength
affects only the overall time scale of the evolution
but not the resulting ensemble of random walks in configuration space.

We therefore simulate the shape {evolution}
 as a random walk of the nuclear shape.
More precisely, we consider a parametrized multi-dimensional family of shapes
suitable for the fission process 
and let $\bold{\chi}=(\chi_1,\chi_2,\dots)$
denote the associated shape parameter.
Let, at some point in the evolution, the nuclear shape be that
defined by a given value of $\bold{\chi}$.
Then, in the course of a brief time $\Delta t$,
the accumulated effect of the nucleons impinging on the surface 
is a stochastic change in the shape, $\Delta\bold{\chi}$,
which can be sampled from the appropriate distribution 
$P(\Delta\bold{\chi};\bold{\chi})$.

Because the potential energy $V(\bold{\chi})$ is given on a fixed lattice
$\{\bold{\chi}_i\}$, 
it is covenient to recast the process as a random walk on that lattice.
This can be accomplished by standard methods.
Due to detailed balance, 
the ratio between the resulting transition probabilities 
for reverse processes equals the Boltzmann factor,
$P(i\to i'):P(i'\to i) = \exp(-\Delta V/T)$,
where $\Delta V\equiv V(\bold{\chi}_{i'})-V(\bold{\chi}_i)$
is the change in the potential energy associated with the shape change
from $\bold{\chi}_i$ to $\bold{\chi}_{i'}$ 
and $T$ is the (local) nuclear temperature.
Such a random walk can readily be simulated by
means of the familiar Metropolis procedure \cite{MetropolisJCP26}.

The above described procedure is merely preliminary and serves to
illustrate the utility of this type of approach.
There is obviously a need to take account of the metric in the shape space,
which is related to the lattice spacings in the employed table 
of deformation energies.
These were chosen in order to achieve typical changes of 
$|\Delta V|\approx1$~MeV and our preliminary studies 
(involving changing the spacings
or introducing a specific metric based on shape overlaps)
suggest that this guiding principle was quite reasonable,
since only relatively extreme modifications have a significant influence
on the results.  A more formal treatment is being developed  \cite{RMS}.

While the Smoluchowski shape trajectory is independent 
of the overall friction strength, the present simplified treatment assumes 
that the friction tensor does not introduce
significant misalignments between the dissipative force 
and the resulting shape change.
Our results are reassuring in this regard 
and a thorough study of this central issue is underway \cite{RMS}.

The five shape parameters used in Ref.\ \cite{MollerPRC79}
are approximately orthogonal, especially near scission and, furthermore, the
mass-asymmetry lattice is equidistant in the fragment mass number $A_{\rm f}$.
Therefore there should be no significant
Jacobian distortion involved in extracting the mass distribution and,
together with the insensitivity to both mass and friction,
this in turn should render the extracted $P(A_{\rm f})$ rather robust,
reflecting primarily the features of the potential-energy surface,
$V(\bold{\chi})$.

For the potential energy, $V(\bold{\chi})$,
we employ tabulated values calculated for the three-quadratic-surface 
shape family \cite{NixNPA130} with the macroscopic-microscopic
finite-range liquid-drop model \cite{MollerPRC79}.
For more than five thousand nuclei,
these tables provide the potential energy of over five million shapes
in terms of five convenient shape parameters.
They are the most comprehensive available and 
have proven to form a good framework for understanding many
important features of fission \cite{MollerNat409}.

The temperature is obtained by the Fermi-gas formula,
$T^2=[E^*-V]/a_A$,
where $E^*$ is the total excitation energy of the nucleus
and $a_A\!=\!A/(8~{\rm MeV})$ is the level-density parameter.
While appropriate for the present explorative study,
this simple approximation may need future refinement, 
such as inclusion of pairing and shell effects.

In order to illustrate the quantitative utility of the dynamical treatment
described above, 
we have used it to calculate the fission fragment mass distribution
for a number of cases of practical interest.
(The calculated mass yields have been reexpressed  as charge yields
by means of a simple scaling, $P(Z_f)=P(A_f)A_0/Z_0$.)

Considering the fission process as a temporal evolution of the nuclear shape,
we combine the original fission theory concepts 
introduced by Bohr \cite{BohrNat143}
with the recognition that the associated statistical distribution 
of nuclear shapes is not established instantly when the nucleus is agitated
(by the absorption of a neutron or a photon) but builds up over time.
Our key assumption is that this equilibration process will terminate
at scission, 
\ie\  when the system finds itself with a shape for which the neck radius
$c_{\rm neck}$  is so small that the nucleus will irreversibly proceed
to separate into two distinct fragments
without any further change in the mass asymmetry. 
We define scission  to occur when $c_{\rm neck}$
has decreased to a specified value $c_0$.
We employ the value $c_0=2.5~{\rm fm}$
but our results tend to be rather insensitive to the precise value.


By discouraging but not prohibiting uphill steps,
the Metropolis sampling method ensures that a 
sufficiently long walk will visit each shape in accordance with its
appropriate statistical weight.
Initially the nuclear shape is close to that of the ground state,
where it resides before being agitated.
The shape will then typically make excursions around this favored shape.
However, every now and then, according to the statistical probability,
the path in deformation space may lead over the fission barrier
and the shape is then more likely to continue towards division
than to revert to a compact shape.

It is thus evident that the random walk will tend to wander around
inside the barrier for many steps before eventually surmounting it.
To speed up the calculation without unduly influencing the final outcome,
we have augmented the potential energy by a bias term,
$V_{\rm bias} = V_0\,Q_0^2/Q^2$,
where $Q$ is the quadrupole moment of the deformed nucleus
and $Q_0$ represents the average ground-state 
quadrupole moment of deformed actinide nuclei.
Thus, in the region of compact shapes, where $Q$ is small,
$V_{\rm bias}$ will encourage increases of $Q$,
while it will have relatively little effect for highly deformed shapes
where the mass division is decided.
The resulting mass yields 
are not sensitive to variations in the bias strength $V_0$,
as long as it remains small (we use $V_0=15$~MeV).

\begin{figure}[t]	
\includegraphics[angle=0,width=3.4in]{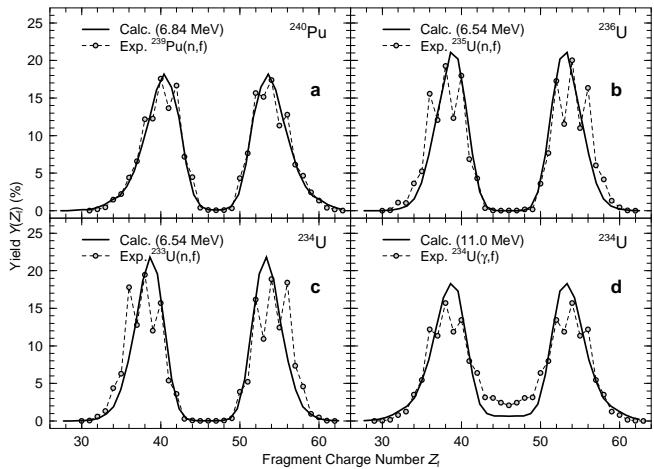}	
\caption{
Calculated and measured charge yields for fission of 
$^{240}$Pu and $^{236,234}$U.
The data in (a--c) are for (n$_{\rm th}$,f) reactions 
leading to $E^*\!\approx6.5$~MeV \cite{ENDF},
while the data in (d) is for ($\gamma$,f) reactions 
leading to $E^*\!\!\approx8-14$~MeV;
they include contamination from fission of $^{233}$U ($\approx\!15\%$) 
and $^{232}$U ($\approx\!5\%$) \cite{SchmidtNPA665};
the corresponding calculation was made for $E^*\!=11$~MeV.
}\label{f:1}
\end{figure}		

Figure~\ref{f:1} shows the calculated charge distributions for 
$^{239}{\rm Pu}$, $^{235,233}{\rm U}{\rm (n_{\rm th},f)}$,
and $^{234}{\rm U}(\gamma,{\rm f})$ 
together with the corresponding experimental data.
(We focus on charge distributions to avoid issues 
related to neutron evaporation.)
The results agree quite well with the data 
which is remarkable because no parameter was adjusted.
The features of the calculated yields are thus determined essentially 
only by the structure of the potential-energy surfaces.
(The odd-even staggering seen in the data is due to pairing
and this effect is not present in the potential-energy surfaces
because existing pairing models treat the fissioning nucleus
as a single system, even near scission.)

The most noticeable discrepancy is an underestimate of the symmetric yield
for the ($\gamma$,f) data (Fig.~\ref{f:1}d).
These were obtained with photons having energies of 8--14~MeV 
\cite{SchmidtNPA665}, while the calculations were made for $E^*=11$~MeV.
Furthermore, 
the experimental data contain contaminations from multi-chance fission.

\begin{figure}[t]	
\includegraphics[angle=0,width=3.4in]{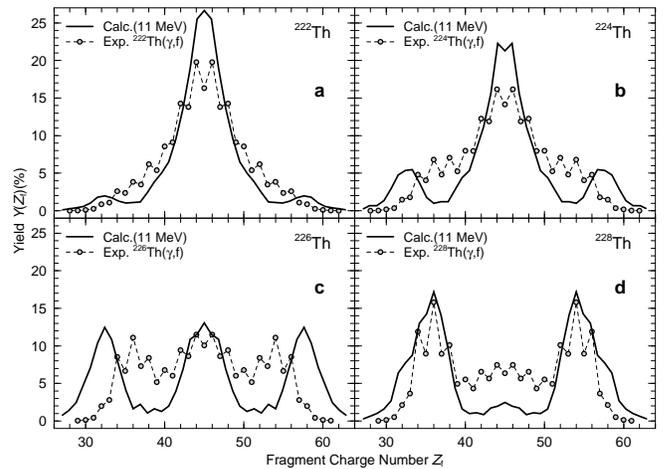}	
\caption{
Calculated charge yields for four even-even thorium isotopes compared
to experimental data \cite{SchmidtNPA665}.
}\label{f:2}
\end{figure}		

We also compare with a sequence of thorium isotopes
which constitute a more challenging test
because their yield curves change qualitatively
from the light to the heavy isotopes.
Figure \ref{f:2} shows calculated and measured charge yields for
$^{222,224,226,228}$Th.
As for $^{234}{\rm U}(\gamma,{\rm f})$ discussed above,
these data \cite{SchmidtNPA665} were obtained with photons 
having a wide energy range and they contain multi-chance contaminations.
Although the differences between calculation and experiment 
are larger than seen in Fig.~\ref{f:1},
the onset of asymmetric fission towards the heavier thorium isotopes
is quite well reproduced.
In fact,  the deviations of the model results from the corresponding data
are smaller than the differences between neighboring experimental results,
which in turn differ by just one occupied neutron orbital. 

The generality of the approach makes it possible to gain new insight
about the fission process from the remaining
differences between calculated results and experimental data. 
For example, 
the energy dependence of the symmetric-valley yield in $^{234}$U (Fig.~1c-d)
can be improved by refining the shape dependence of the temperature 
and an alternate shape dependence of the Wigner term in the potential energy
\cite{MollerNPA492} moves the asymmetric peaks in Fig.~2c 
into agreement with experiment.

The case of $^{222}$Th (Fig.~\ref{f:2}a) is particularly instructive,
because the calculation yields a symmetric mass distribution
even though the nuclear shape at the saddle point of the
potential energy surface is reflection asymmetric
with a mass ratio of 129:93.
This remarkable finding implies that either the path from the isomeric minimum 
(which is located at a symmetric shape) to scission
does not cross the potential barrier in the most favorable region
({\em i.e.}\ near the saddle point)
or the shape distribution reverts from asymmetric to symmetric 
during the descent from saddle to scission.
In either case, the result invalidates the hypothesis
(see {\em e.g.}\ Refs.\ 
\cite{MollerNPA192,BolsterliPRC5,BenlliureNPA628,SchmidtNPA693})
that the character of the mass distribution,
whether symmetric or asymmetric, is determined by the saddle shape.
Rather, our result suggests that the fragment mass distribution 
is determined by the relatively complicated structure of the potential-energy
landscape between the isomeric minimum and scission.
Therefore any plausible model of the mass yields must take this into account. 

It is interesting to compare the calculated mass yields
with the result of statistical scission models \cite{FongPR102,WilkinsPRC14}.
We have found that while a statistical sampling of our scission shapes
often (but not always) lead to reasonable agreement with the data,
it is never as good as the results of the random walk,
suggesting that the shape evolution in the pre-scission landscape generally 
plays an important role for the resulting mass distribution \cite{RMS}.


In summary, we have presented a novel treatment of the shape dynamics
in moderately excited nuclei
and we have illustrated its practical and quantitative utility
by using it to calculate fission fragment yields
for several cases that have been studied experimentally,
including some particularly challenging ones.
Relative to previously employed methods, 
the present approach represents a significant advance
with regard to predictive power.
(We have here concentrated on nuclei with 5-20~MeV excitation
but we plan to explore extensions to both higher and lower energies,
including spontaneous fission.)

Taking explicit account of the equilibration process,
our treatment extends in a natural way the compound nucleus concept
invoked in 1939 to describe many aspects of
the newly discovered fission phenomenon.
It builds directly on the general picture of low-energy 
nuclear dynamics as being dominated by the dissipative interaction between 
the evolving surface and individual nucleons.
This mechanism causes the nuclear shape dynamics
to resemble Brownian motion
and the present dynamical treatment is the first
to treat the resulting stochastic shape evolution in the 
Smoluchowski limit where the inertial mass is immaterial.
A particularly attractive feature of the approach is its generality:
once the potential energy has been calculated as a function of deformation,
for a sufficiently rich class of mononuclear shapes, 
the dynamics can readily be studied.
Studies are well underway to clarify the importance of
the shape metric, the friction tensor,
and pairing and shell effects in the entropy \cite{RMS}.

We have here concentrated on applications of this treatment
to the calculation of fission fragment mass distributions
for which a variety of data is available.
Importantly, only a single new parameter is required for this purpose,
namely the critical neck radius characterizing a scission shape,
and the mass yields are rather insensitive to its specific value.
This degree of robustness gives the method unprecedented predictive power
with regard to fission-fragment mass distributions.
In particular, it can be readily employed in regions of the nuclear chart
that are of special astrophysical interest and it may, for example,
help to clarify the importance of fission recycling for the $r$-process
\cite{Seeger,BreunPRC77}.\\

\noindent{\bf Acknowledgments}\\
We are grateful to K.-H.\ Schmidt for providing computer-readable files
of the data in Ref.\ \cite{SchmidtNPA665} and to
L.~Bonneau, H.~Goutte, D.C.~Hoffman, A.~Iwamoto,
A.J.~Sierk, and R.~Vogt for helpful discussions; 
T.~Watanabe kindly extracted the (n,f) data from the ENDF/B-VII.0 data base.

This work was supported by the Director, Office of Energy Research,
Office of High Energy and Nuclear Physics,
Nuclear Physics Division of the DOE 
under Contract No.\ DE-AC02-05CH11231 (JR)
and by the National Nuclear Security Administration 
of the U.S.\ Department of
Energy at Los Alamos National Laboratory 
under Contract No.\ DE-AC52-06NA25396 (PM).\\

		

			\end{document}